\documentstyle[prl,aps,multicol,epsf,rotate]{revtex}
\def\Et{{\widetilde E}}
\def\At{{\widetilde A}}
\def\Tt{{\widetilde T}}
\begin{document}
\draft
\preprint{XXXX}
\title{Logarithmic Correlations in Quenched Random Magnets and Polymers}
\author{John Cardy}
\address{University of Oxford, Department of Physics -- Theoretical
         Physics, 1 Keble Road, Oxford OX1 3NP, U.K. \\
         and All Souls College, Oxford.}
%
%
\maketitle
\begin{abstract}
It is argued that logarithmic factors multiplying power law behavior
are to be expected in the scaling behavior at or near non-mean field
critical points of systems with short-range interactions
described theoretically by any kind of $n\to0$ limit, in which
the effective free energy vanishes. Explicit examples are given for 
quenched random ferromagnets, polymer statistics and percolation, but the 
phenomenon is quite general. 
\end{abstract}
\begin{multicols}{2}
Logarithmic corrections to power law behavior
at or near a critical point are normally associated
with the presence of operators which are marginally irrelevant under the
renormalisation group. The amplitudes of such logarithmic terms vanish 
if the critical theory happens 
to lie exactly at the fixed point. However, it has been pointed out
\cite{Gur} that the structure of two-dimensional
conformal field theory, which describes the fixed point in a wide range
of isotropic systems with short-range interactions, allows so-called 
\em logarithmic \em operators in the fixed point theory itself,
whose operator product expansions and correlation functions contain
logarithms as well as pure power behavior. 

Such operators should not occur in unitary conformal field theories,
such as correspond to
pure critical systems with positive Boltzmann weights, for the
following reasons.
Any real local density $\Phi(r)$ may always be expanded 
in a series of scaling operators $\Phi(r)=\sum_ia_i\phi_i(r)$, 
so that its two-point
correlation function in ${\bf R}^d$ takes the form
\begin{equation}
\langle\Phi(r)\Phi(0)\rangle\sim\sum_{ij}{A_{ij}\over r^{x_i+x_j}}
\label{decomp}
\end{equation}
where the scaling dimensions $x_i$ are the eigenvalues of the
dilatation operator $\cal D$. The off-diagonal terms with $x_i\not=x_j$ vanish
by conformal symmetry. 
In addition, the diagonal coefficients $A_{ii}$ are all non-negative
as a consequence of reflection positivity.
Thus, in such a theory, only pure powers, or sums of them, 
should ever appear \cite{footnotecont}. 
However, in non-unitary conformal field theories, 
it is possible for some of the $A_{ii}$ to be
negative, in which case cancellations may occur in (\ref{decomp}). In
particular, if two scaling dimensions $x_i$ and $x_j$ become
degenerate in such a way that $A_{ii}\sim-A_{jj}\to\infty$ with 
$A_{ii}(x_i-x_j)$ remaining finite, the leading
terms will cancel leaving a logarithmic term proportional to
$r^{-2x_i}\ln r$. 

The main point of this paper is that such logarithmic terms
should always occur in models which are described by non-trivial fixed
point
theories with a trivial partition function $Z=1$, in an arbitrary
number of dimensions. These include all
sorts of quenched random systems including percolation,
as well as problems such as
self-avoiding random walks describing polymer statistics. 
Such logarithmic operators have already
been identified in some systems \cite{dis} which may be analysed using
supersymmetry to perform the quenched average, and there has been
considerable work \cite{gen} on the general structure of such logarithmic
CFTs. We shall use instead the replica formalism, which has the advantage of
being more generally applicable, as well as providing a simple way of
seeing how the logarithmic terms appear in the limit. 
It will be seen that logarithms appear as the leading contributions
in some quite simple two-point
correlation functions, as well as completely generally as subleading
terms in higher order ones.

The analysis depends on some very simple properties of conformal field
theory which are valid in all dimensions, not only $d=2$. That is, they
should hold for all critical theories of the above types as long as $d$
is below the upper critical dimension.  A crucial role is also
played by the symmetry group of the theory, be it the permutation group
$S_n$ of the replicas, or $O(n)$ for polymers. Rather than work in complete
generality we shall give two illustrative examples which are moreover
physically relevant. 

Consider first a quenched random ferromagnet, for example an Ising model
with random bond disorder. Such systems may be described generically in
the continuum limit by a reduced hamiltonian 
$H=H^0+\int m(r)E(r)d^d\!r$, where $H^0$ describes the pure system and
$m(r)$ is a quenched random variable coupled to the local energy density
$E(r)$. The disorder is supposed to have short-range correlations only,
so that $\overline{m(r)}=0$ and 
$\overline{m(r)m(r')}=g\delta_{rr'}$. Introducing replicas
$a=1,\ldots,n$ in the standard way and averaging over the disorder gives
the replica hamiltonian
\begin{displaymath}
H_R\equiv\sum_aH_a^0-g\int\sum_{a\not=b}E_a(r)E_b(r)d^d\!r
\end{displaymath}
where the higher cumulants have been dropped on the basis that they are
irrelevant in the RG sense. The terms with $a=b$ are discarded because
they can be further reduced using the operator product expansion (OPE)
in the pure theory which takes the form
$E_a\cdot E_a=1+b_0E_a+\cdots$.
The dimension of $g$ is $y_g=d-2x_E^0$, and it is relevant at the pure fixed
point if $d>2x_E^0$. This is equivalent to the well-known Harris
criterion $d\nu^0<2$ \cite{Harris}. If $y_g$ is small, it is possible to develop
a perturbative RG in powers of this variable and to locate a random
fixed point to which critical systems with sufficiently weak randomness
should flow \cite{Ludwig,JCbook}. 
The shift in the scaling dimension $x_\phi$ of some
scaling operator $\phi$ from the pure to the random fixed point is given by
\cite{JCbook}
\begin{displaymath}
x_\phi=x_\phi^0+(2b_\phi/b)y_g+O(y_g^2)
\end{displaymath}
where $b$ and $b_\phi$ are the coefficients in the OPEs
$\Phi\cdot\Phi=1+b\Phi+\cdots$ and
$\phi\cdot\Phi=b_\phi\phi+\cdots$, and $\Phi$ is the perturbing
operator, in this case $\sum_{a\not=b}E_aE_b$, which, using the above
OPE of the pure theory, gives $b=4(n-2)+2b_0^2$.

Now consider the case where $\phi$ is itself the replicated
energy density $E_a$. The main point is that all these $n$ operators are
degenerate at the pure fixed point, and, to carry out the perturbative
expansion, they should be decomposed into irreducible representations of
the replica permutation group $S_n$. In this case this is
straightforward: the combination $E\equiv\sum_aE_a$ is a singlet, leaving
$\Et_a\equiv E_a-(1/n)\sum_bE_b$, with $\sum_a\Et_a=0$,
transforming according to
an $(n-1)$-dimensional representation. The OPE coefficients are easily
found: $b_E=2(n-1)$ and $b_\Et=-2$, so that, close to $n=0$,
$x_E(n)=x_E^0+\frac12(1-n)y_g+O(y_g^2)$ and
$x_\Et(n)=x_E^0+\frac12y_g+O(y_g^2)$.
Note that the two scaling dimensions agree at $n=0$. This is in fact
true to all orders, and is a consequence of the fact \cite{JCmod}
that the partition
function of any CFT on ${\bf S}_1\times{\bf S}_{d-1}$ takes the form
$Z=\sum_iq^{x_i}$, where $q$ is a modular parameter. Since $Z=1$ at $n=0$,
all scaling operators must arrange themselves to have multiplicities
proportional to $n$ as $n\to0$. Thus the singlet operator $E$ becomes
degenerate with the $(n-1)$ operators $\Et_a$. Assume that the
difference between the scaling dimensions vanishes linearly in $n$, as
it does to first order.

The two-point functions therefore take the form, for $n\not=0$,
\begin{eqnarray}
\langle E(r)E(0)\rangle&=&n\left(\langle E_1E_1\rangle
+(n-1)\langle E_1E_2\rangle\right)\nonumber\\
&\sim&nA(n)r^{-2x_E(n)}\label{EE}\\
\langle\Et_a(r)\Et_a(0)\rangle&=&(1-1/n)\left(\langle E_1E_1\rangle
-\langle E_1E_2\rangle\right)\nonumber\\
&\sim&(1-1/n)\widetilde A(n)r^{-2x_\Et(n)}\label{EtEt}
\end{eqnarray}
where replica symmetry has been used in the first equalities.
The amplitudes $A(n)$ and $\widetilde A(n)$ have finite limits at $n=0$
in the pure theory and therefore should remain so at the random fixed
point. For consistency  between (\ref{EE}) and (\ref{EtEt})
$A(0)=\widetilde A(0)$. Subtracting
the two expressions and letting $n\to0$,
\begin{eqnarray*}
\overline{\langle E(r)\rangle\langle E(0)\rangle}&=&
\lim_{n\to0}\langle E_1(r)E_2(r)\rangle\nonumber\\
&\sim&2A(0)x_E'(0)\ln r/r^{-2x_E(0)}
\end{eqnarray*}
The left hand side of this equation gives the physical meaning of
the replica expression $\langle E_1E_2\rangle$: it is the quenched
average of the \em disconnected \em part of the energy-energy
correlation function. Note that the connected piece
$\overline{\langle EE\rangle}-\overline{\langle E\rangle\langle
E\rangle}=\langle E_1E_1\rangle-\langle E_1E_2\rangle$ is given by the
$n\to0$ limit of both equations above and contains no logarithm.
Although the amplitude $A(0)$ is non-universal, the
ratio of the quenched average of the disconnected piece to that of the
connected piece is asymptotically proportional to $\ln r$ with a \em
universal \em amplitude. 

Such a phenomenon does not occur for the correlation functions of the
local magnetisation $\sigma$ \cite{footnoteRFIM}. 
Taking the example of an Ising model, the
replicated hamiltonian is invariant under independent $Z_2$ transformations
$\sigma_a\to-\sigma_a$ in each replica, as well as replica permutations.
Under this larger symmetry, the operators $\{\sigma_a\}$ transform
irreducibly, so that $x_\sigma(n)=x_{\tilde\sigma}(n)$ for all $n$.
However, as argued below, logarithms do appear in higher-point
correlations of $\sigma$. Logarithms also appear in certain two-point
functions of composite operators $E_{a_1}E_{a_2}\ldots E_{a_p}$, which
correspond to higher moments $\overline{\langle EE\rangle^p}$ of the
two-point function (and which exhibit multiscaling). 
These arise \cite{TDJC} from a similar mechanism to the above:
the singlet operator $(\sum_bE_b)^p$ becomes degenerate with the
$(n-1)$ independent operators $(\sum_bE_b)^{p-1}(E_a-(1/n)\sum_bE_b)$,
but only exactly at $n=0$.
Note also that it has been observed \cite{CKT} that logarithmic
operators whose 2-point functions behave like 
$\langle D(r)D(0)\rangle\sim r^{-2x}(\ln r + {\rm const.})$ are always
accompanied by another operator $C$ such that $\langle
C(r)D(0)\rangle\sim
r^{-2x}$ and $\langle C(r)C(0)\rangle=0$. In this example, $C=E$ and
$D=E_a$, in the limit $n=0$, so that $D$ is a reducible operator under
the replica symmetry.

As a second example, consider self-avoiding walks, which are a model for
polymer statistics, and are described \cite{dG}
by the $n\to0$ limit of an $n$-component
field theory with a hamiltonian with $O(n)$ symmetry
$\int\big(\sum_a((\nabla\phi_a)^2+m^2\phi_a^2)+g\sum_{ab}\phi_a^2
\phi_b^2\big)d^d\!r$. Consider the composite operators of the form
$:\!\phi_a\phi_b\!:$ (the normal ordering subtracts the disconnected
part $\langle\phi_a\phi_b\rangle$). 
These may be decomposed under $O(n)$ into a singlet
$E\equiv\sum_a:\!\phi_a^2\!:$ and a symmetric traceless part 
$\Et_{ab}\equiv:\!\phi_a\phi_b\!:-(1/n)\delta_{ab}\sum_c:\!\phi_c^2\!:$, with
$(n-1)(n+2)/2$ independent components. In general, these two irreducible
sets carry different scaling dimensions, as can checked in perturbation
theory in $g$: at one loop, the relevant OPE coefficients are
$b_E=8+4n$ and $b_\Et=8$. Once again, they become degenerate at $n=0$
so that $Z=1$ \cite{footnote2d}. Their two-point functions have the form
\begin{eqnarray*}
\langle E(r)E(0)\rangle&=&2n A(n)r^{-2x_E(n)}\\
\langle\Et_{ab}\Et_{cd}(0)\rangle&=&
(\delta_{ac}\delta_{bd}+\delta_{bc}\delta_{ad}-
{\textstyle{2\over n}\displaystyle}\delta_{ab}\delta_{cd})
\At(n)r^{-2x_\Et(n)}
\end{eqnarray*}
where the index structure is fixed by group theory and the amplitudes
$A(n)$ and $\At(n)$ are finite at $n=0$. Now consider the
correlation function\break 
$\langle:\!\phi_a(r)\phi_b(r)\!::\!\phi_a(0)\phi_b(0)\!:\rangle$
and express this in terms of $\langle EE\rangle$ and
$\langle\Et\Et\rangle$. There is then a singular
term proportional to $(\delta_{ab}\delta_{cd}/n)\big(A(n)r^{-2x_E(n)}-
\At(n)r^{-2x_\Et(n)}\big)$, which gives rise to a logarithm at $n=0$.
A similar argument may be made away from the critical point, where 
the two-point function of each scaling operator should scale as
$\xi^{-2x_i(n)}f_i(r/\xi,n)$ where $\xi$ is the correlation length. This
implies that the generalized susceptibilities $\chi_E\equiv
\sum_r\langle E(r_1)E(r_2)\rangle$ and 
$\chi_\Et\equiv\sum_r\langle\Et(r_1)\Et(r_2)\rangle$
should behave as $C_i(n)\xi^{d-2x_i(n)}$, where $i\in\{E,\Et\}$
and $C_E(0)=C_\Et(0)$.

The physical interpretation of this is as follows. The term proportional
to $\delta_{ab}\delta_{cd}$ in the full correlation function
counts all pairs of mutually avoiding
self-avoiding loops (polygons on a lattice) one of which goes through
$r_1$, and the other through $r_2$. The normal ordering 
subtracts off the disconnected piece which corresponds to the 
product of the independent sums over such pairs of loops, in which they
are allowed to intersect. Therefore the remaining connected part, to
which the above is the leading contribution, corresponds to all
configurations where the two loops intersect at least once.
Denoting by $c_N$ the number of such pairs of loops, of total length $N$,
per lattice site, the mapping to the $O(n)$ model is such that
its generating function $\sum_NN^2c_Ny^N$ is proportional to $\lim_{n\to0}
(\chi_E-\chi_\Et)/n$, with $\xi\propto(y_c-y)^{-\nu}$. 
This implies that $c_N\sim N^{(d-2x_E)\nu-3}\mu^N\ln N=
N^{\alpha-3}\mu^N\ln N$, where $\mu=y_c^{-1}$. For monodisperse loops,
there is an additional factor of $N^{-1}$.
This agrees with an alternative method: the free energy per site $f(n)$
of the $O(n)$ model scales as $B(n)(x_c-x)^{2-\alpha(n)}$, where
$B(0)=0$. Since each loop appears with fugacity $n$,
the second derivative $\partial f/\partial n|_{n=0}$ counts pairs of
intersecting loops, and its leading singularity
\cite{footnotesmallloops} is of the form
$B'(0)\alpha'(0)(x_c-x)^{2-\alpha(0)}\ln(x_c-x)$, consistent with
the above result.

Although we have given two examples of how logarithmic factors occur in
two-point functions, it is worth pointing out that they are ubiquitous
in four- and higher-point functions
in fixed point theories with $Z=1$. This is because conformal invariance
fixes the coefficient of the stress tensor $T$ in the OPE of an operator
$\phi$ with itself to have the form 
$\phi\cdot\phi=a_\phi({\bf 1}+(x_\phi/c)T+\cdots)$, where all
the co-ordinate dependence and the complicated index structure has been
suppressed \cite{JCani}. 
Here $a_\phi$ defines the normalisation of $\phi$ and $c$
is the central charge, defined in arbitrary dimension in terms of the
two-point function $\langle T(r)T(0)\rangle\propto c/r^d$, where again
the index structure has been suppressed. Substituting these terms of
the OPE into the 4-point function 
$\langle\phi(r_1)\phi(r_2)\phi(r_3)\phi(r_4)\rangle$
in the limit where $\eta\equiv r_{12}r_{34}/r_{13}r_{24}\ll1$ gives a
term proportional to $(r_{12}r_{34})^{-2x_\phi}(a_\phi^2/c)\eta^d$.
In a theory such as those discussed above, with $c=0$, this raises a
potential
difficulty \cite{Gur2}. The resolution is that, in general, there are other
operators whose scaling dimensions become degenerate with that of $T$ as
$n\to0$. For the quenched random ferromagnet, these are the operators
$\Tt_a\equiv T_a-(1/n)\sum_bT_b$, constructed in a similar way to the
$\Et_a$ above. For the $O(n)$ model, they are traceless part of
$\Tt^{\mu\nu}_{ab}\equiv\partial^\mu\phi_a\partial^\nu\phi_b+\cdots$, 
whose trace over $a=b$
gives the stress tensor $T^{\mu\nu}$. These operators give rise to
logarithmic terms of the form $\eta^d\ln\eta$ in the four-point
functions at $n=0$. 

However, the paradox may be resolved in another way. The physical
correlation functions may be such that $a_\phi\to0$ as $c\to0$. This in
fact happens in the $q$-state Potts model where all connected
correlations vanish linearly at $q=1$, so that $a_\phi\propto c$. The
connectivities of the percolation problem are given by the derivatives
with respect to $q$ at $q=1$, and are finite. But in this case there are
no logarithmic terms of the above form \cite{footnotettbar}.

It would be very interesting to understand how the $(n-1)$ bosonic
partners of the
stress tensor like $\Tt$ become its fermionic partners 
in those cases where the quenched average can be performed
using supersymmetry \cite{Gur3}. 

We have argued that logarithmic factors multiplying power law
singularities are ubiquitous in  critical systems described by effective
theories with $Z=1$. In general they hide themselves in non-leading
corrections to higher-point correlations, but there are examples
where they appear in two-point correlations and susceptibilities. 
Although all conclusions are for general $d$, it is
possible to check some of them in $d=2$ for percolation and the $O(n)$
model \cite{JCinprog}, where many of the critical four-point functions are known
exactly. 

The author thanks V.~Gurarie, I.~Kogan and H.~Saleur for useful
discussions, and the Aspen Center for Physics for its hospitality.
This research was supported in part by the Engineering and 
Physical Sciences Research Council under Grant GR/J78327.

\end{multicols}


\begin{references}
\bibitem{footnotecont} Logarithmic factors could also arise if the
spectrum of $\cal D$ were continuous, but these would in general not be
positive integral powers.
\bibitem{footnotesmallloops} In fact there is a more important term 
$O(N^{\alpha-2}\mu^N)$ which mathematically arises form the fact that
$y_c$ depends on $n$, and physically from the contribution when one of
the loops is small. This is removed by considering loops of equal
length.
\bibitem{footnotettbar} However, at least in $d=2$, logarithms do arise
at a higher level because the coefficient of $T\overline T$ in the OPE
is $O(1/c^2)$.
\end{references}
\end{document}